\documentclass[12pt,preprint]{aastex}

\usepackage{graphicx}

\slugcomment{Revised 2013-June-11}

\shorttitle{Phaethon Tail}
\shortauthors{Jewitt, Li and Agarwal}

\begin{document}

\title{The Dust Tail of Asteroid (3200) Phaethon
}
\author{David Jewitt$^{1,2}$, Jing Li$^1$ and
Jessica Agarwal$^3$
}
\affil{$^1$Department of Earth and Space Sciences,
University of California at Los Angeles, \\
595 Charles Young Drive East, 
Los Angeles, CA 90095-1567\\
$^2$Department of Physics and Astronomy,
University of California at Los Angeles, \\
430 Portola Plaza, Box 951547,
Los Angeles, CA 90095-1547\\
$^3$ Max Planck Institute for Solar System Research, Max-Planck-Str. 2, 37191 Katlenburg-Lindau, Germany
}

\email{jewitt@ucla.edu}

\begin{abstract}
We report the discovery of a comet-like tail on asteroid (3200) Phaethon when imaged at optical wavelengths near perihelion. In both 2009 and 2012,  the tail appears $\gtrsim$350\arcsec ~(2.5$\times$10$^8$ m) in length and extends approximately in the projected anti-solar direction.  We interpret the tail as being caused by dust particles accelerated by solar radiation pressure.  The sudden appearance and the morphology of the tail indicate that the dust particles are small, with an effective radius $\sim$1 $\mu$m and a combined mass $\sim$3$\times$10$^5$ kg.  These particles are likely products of thermal fracture and/or desiccation cracking under the very high surface temperatures ($\sim$1000 K) experienced by Phaethon at perihelion.    The existence of the tail confirms earlier inferences about activity in this body based on the detection of anomalous brightening.  Phaethon, the presumed source of the Geminid meteoroids, is still active.
\end{abstract}


\section{Introduction}
Asteroid (3200) Phaethon is a $\sim$5 km diameter body dynamically associated with the Geminid meteoroid stream \citep{1983IAUC.3881....1W} and with several kilometer-scale asteroids collectively known as the Phaethon-Geminid complex (PGC; Ohtsuka et al.~2009, Kasuga 2009).  Most meteoroid streams have cometary parents (Jenniskens 2008) from which mass loss is driven by the sublimation of near-surface ice. However, Phaethon is dynamically an asteroid (semimajor axis 1.271 AU, eccentricity 0.89, inclination 22.2, Tisserand parameter relative to Jupiter, $T_J$ = 4.54), raising questions about the mechanisms by which it loses mass.  

The formation of the PGC could be ancient \citep{2009PASJ...61.1375O}, but the short dynamical lifetime of the Geminid meteoroid stream \citep[$\sim$10$^3$ yr;][]{1989A&A...225..533G,2007MNRAS.375.1371R}, opens the possibility that Phaethon might still be active.  However, attempts over two decades to detect activity in Phaethon have proved negative \citep{1984Icar...59..296C,1996Icar..119..173C,2005ApJ...624.1093H,2008Icar..194..843W}.  The first evidence for continuing activity was obtained only recently.  Photometry in both 2009 \citep{2010AJ....140.1519J} and 2012 \citep{2013AJ....145..154L} showed anomalous perihelion brightening, in which the apparent brightness increased suddenly at large phase angles, opposite to the fading trend expected from the phase function of a solid body.  Numerous mechanisms (thermal emission, glints, fluorescence stimulated by the impact of the solar wind, sublimation of embedded ice, prompt emission from forbidden transitions in atomic oxygen) were considered and found incapable of producing the anomalous brightening (Li and Jewitt 2013).  In particular, near-surface water ice is thermodynamically unstable on Phaethon as a result of its high surface temperature.  Deeply-buried water ice would be thermally insulated and phase-lagged from the surface heat, leaving no explanation for the coincidence between activity and perihelion \citep{2010AJ....140.1519J} and \citep{2013AJ....145..154L}. Instead, the release of dust from the nucleus is able to explain the data in a plausible way.  The process responsible for forming and ejecting the dust is presumed to relate to the high ($\sim$1000 K) temperatures attained by the surface of Phaethon when at perihelion ($q$ = 0.14 AU).  Thermal fracture and cracking due to desiccation shrinkage of hydrated silicates are two processes capable of both producing the dust and ejecting it from the surface \citep{2010AJ....140.1519J,2012AJ....143...66J,2013AJ....145..154L}.  

Observations  at perihelion are extremely challenging, because the solar elongation then is small ($<$8\degr) and Phaethon must be viewed against the bright, structured and changing background of the solar corona. Here, we use data from the NASA-STEREO coronal imaging spacecraft to perform a search for spatially-resolved evidence of activity at perihelion. 

\section{Observations} 

We used the Heliospheric Imagers (HI) from the Sun Earth Connection Coronal and Heliospheric Investigation (SECCHI) package \citep{2008SSRv..136...67H,2009SoPh..254..387E} on the NASA STEREO spacecraft.  Our observations exclusively employed the STEREO-A HI-1 camera, having a field center offset from the solar center by 14$^\circ$ and with a square field of view $20^\circ$ in width.  The 2048$\times$2048 pixel charge-coupled device (CCD) detectors are binned 2$\times$2 before transmission to Earth.  The resulting angular size of each pixel is 70\arcsec.  

Each 1024$\times$1024 pixel-image is compiled from a set of 30 integrations each of 40 s, and taken at 1 minute intervals. A single downloaded image therefore has an effective exposure time of 1200 s (20 minutes). One such image is obtained every 40 minutes.  The onboard combination of multiple short-exposure images permits the rejection of cosmic rays and other artifacts, and avoids saturation of the background corona that would otherwise occur owing to the large pixels in HI-1.  The quantum efficiency of the camera is practically uniform across the  6300 to 7300 \AA~wavelength passband \citep{2009SoPh..254..387E}.

We searched for extended emission in the HI-1 images used in our earlier work \citep{2013AJ....145..154L}.
Our procedure removed large angular scale structures in the coronal background, but left small scale and rapidly varying features, as well as background stars.  
We used NASA's HORIZONS software to compute the position of Phaethon as seen from the STEREO spacecraft and calculated the expected location on the CCD in pixel coordinates.

When displayed in rapid succession as a movie, the images from 2009 hint at the presence of a tail on Phaethon, but fluctuations in the surface brightness of the coronal and sky background from image to image are much larger than the surface brightness of the tail itself.  Simple median stacks show that the tail appears concurrently with the anomalous brightening but is otherwise absent.  Unlike the background fluctuations, the statistical significance of the tail grows as more images are combined.   To test the possibility that the tail might be an artifact produced by only a fraction of the data, we separately combined subsets of the images (0.5 to 0.8 days at a time).  The subsets all showed the tail but, as expected, at lower significance owing to the smaller number of images in the subsets.   

\begin{figure}
\epsscale{1.}
\begin{center}
\plotone{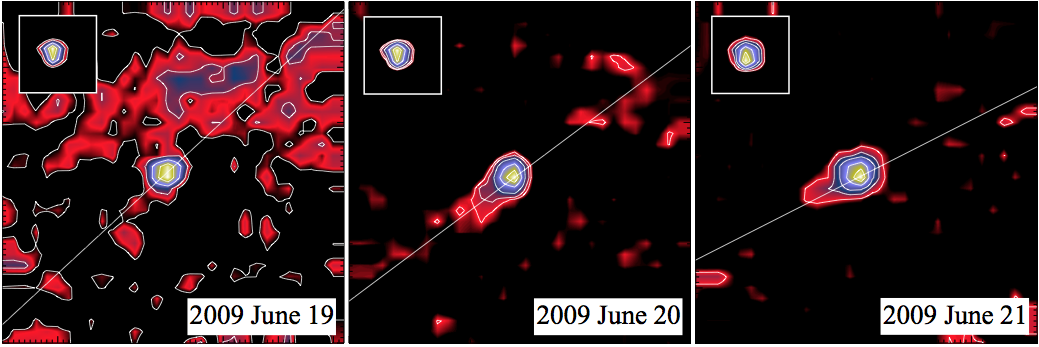}
\caption{Composite images of (3200) Phaethon in 2009 compared with the projected sun-comet line (white).  The Sun is to the upper right in each panel. Insets are 490\arcsec~square and show field stars near to Phaethon to demonstrate the point spread function of the data.  Each panel has North to the top, East to the left and shows the median of $\sim$30 images taken over a 1 day period starting at the times listed in Table 1. Anomalous brightening peaks were reported on June 20 and 21 in Jewitt and Li (2010).    \label{img2009}
} 
\end{center} 
\end{figure}

Next, since the projected antisolar direction, $\theta_{\odot}$, changes rapidly in the period of interest, we re-combined the images including a correction for the changing $\theta_{\odot}$.  We removed field stars from the images by hand prior to computing the median of an image stack.    The resulting images improve the apparent brightness of the tail and show that it is aligned with the projected sun-Phaethon line (Figure \ref{img2009}).  To test the possibility that the improvement in the de-rotated images might be a result of chance in noisy data, we repeated the procedure but for a wide range of unphysical rotations.  In these unphysical image combinations, the Phaethon tail became washed out or invisible, as expected if the tail is real.   To test the possibility that the tail might be caused by a peculiar asymmetry or astigmatism in the images from the HI-1 camera, we examined the images of comparably bright field stars located close to the path of Phaethon in the period of interest.  The stars showed no asymmetry and no evidence for a Phaethon-like tail (see the inset images of field stars in each panel of the 2009 data in Fig. \ref{img2009}).  Lastly, we note that no tail was detected in the image composite having start-time UT 2009 June 19d 06h 49m (i.e.~one day pre-perihelion, left panel of Figure \ref{img2009}).  The tail was detected only on the two subsequent days, coinciding with the anomalous brightening reported in \citet{2010AJ....140.1519J} and \citet{2013AJ....145..154L}. There are no useful later data from the STEREO spacecraft.

\begin{figure}
\epsscale{1.}
\begin{center}
\plotone{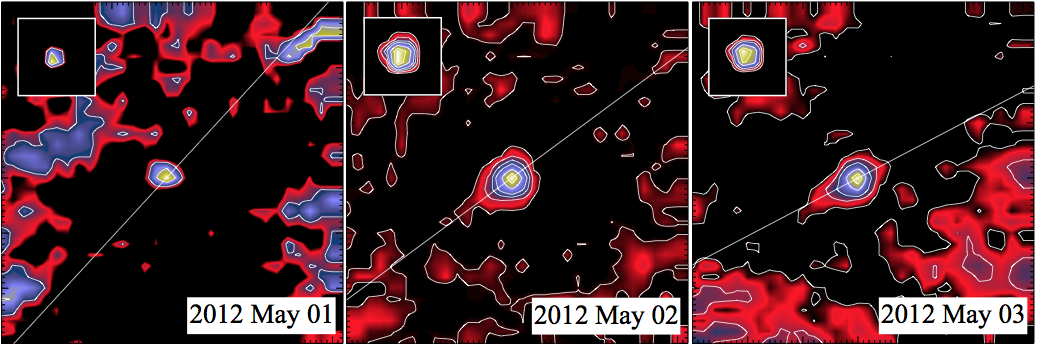}
\caption{Same as Figure (\ref{img2009}) but for data from 2012 (see Table \ref{positionangles}).  Anomalous brightening peaks were detected on May 02 and 03 in Li and Jewitt (2013). \label{img2012}
} 
\end{center} 
\end{figure}

\begin{figure}
\epsscale{1.0}
\begin{center}
\plotone{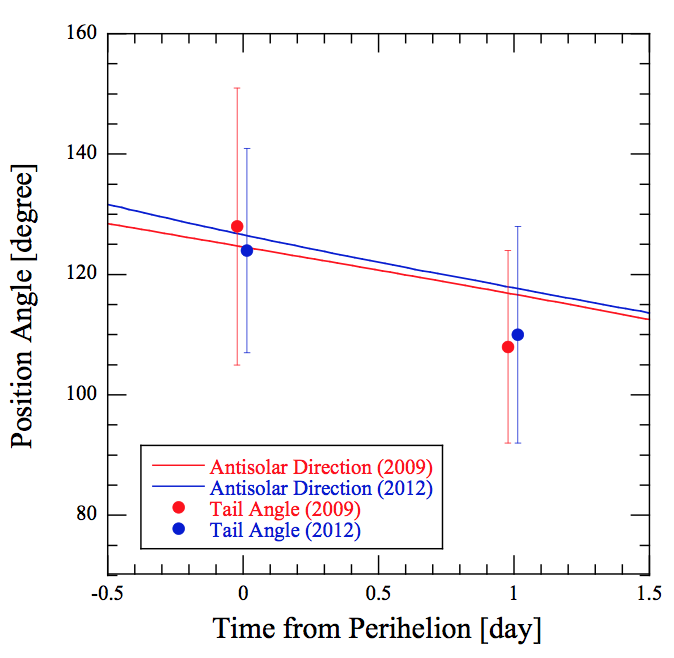}
\caption{Measured position angle of the tail  in 2009 (red circles) and 2012 (blue circles), compared with the projected antisolar direction (red and blue lines).   Error bars on the position angle measurements show the estimated 1-$\sigma$ uncertainties from composite images formed by averaging the data over 1-day intervals, as described in the text.    \label{tail_pa}
} 
\end{center} 
\end{figure}

The entire procedure was repeated using the data from 2012, with the same result (Figure \ref{img2012}).  In both years Phaethon shows a faint, approximately antisolar tail that becomes brighter when the blurring effects of differential image rotation in the image sequence are correctly removed and fainter, to the point of disappearing, when they are not.  The tail appears only on the two days for which Phaethon showed anomalous brightening \citep{2013AJ....145..154L}. We measured $\theta_{Ph}$, the position angle of the Phaethon tail and present the results in Table (\ref{positionangles}), along with the geometric circumstances of Phaethon in each year.  The uncertainties on $\theta_{Ph}$, determined from azimuthal surface brightness profiles centered on Phaethon, reflect the large pixel scale, the faintness of Phaethon's tail and the complexity of the sky background.  The measured $\theta_{Ph}$ are also plotted in Figure (\ref{tail_pa}), where it may be seen that  $\theta_{Ph}$ and $\theta_{\odot}$ are identical within the uncertainties of measurement.  

\section{Discussion}

Thermal emission, specular reflection ``glints'', fluorescent excitation by the solar wind and prompt emission from the excited $^1$D level of [OI] (produced by the photo-destruction of water) were all considered and rejected as sources of the anomalous perihelion brightening \citep{2010AJ....140.1519J,2013AJ....145..154L}.  The first three would all produce brightening only of the nucleus (i.e.~the central pixel in our data) and hence are additionally inconsistent with the detection of a resolved tail.  Sodium is depleted in Geminid meteors \citep{2009EM&P..105..321K} and might be baked-out from the nucleus of Phaethon.  The Na D-lines at $\sim$5890\AA, however, fall outside the 6300 to 7300 \AA~passband of the HI-1 camera and so cannot contribute to the tail.  Prompt emission from  the forbidden lines of oxygen at 6300\AA~and 6363\AA~falls within the instrumental passband but would require a large production rate of $\sim$10$^{30}$ s$^{-1}$ to match the observed brightening \citep{2013AJ....145..154L}.  Furthermore, the photodissociation lifetime of water in sunlight at 0.14 AU is only $\tau_{d}$ = 0.5 hr \citep{1992Ap&SS.195....1H}.  It is unlikely that water molecules could travel the length of the tail in such a short time.   

Our preferred interpretation is that the tail of Phaethon is a dust tail.  The key observables are the time of appearance of the tail, the length of the tail and the position angle of the tail in the plane of the sky (Table \ref{positionangles}).  We computed the running median of $\sim$30 images having a range of start-times around perihelion.  In both years, the emergence of the tail corresponded with times of perihelia (2009 June 20 07:22 and 2012 May 02 07:49). The length of the tail in the plane of the sky was estimated at $\ell \sim$ 250,000 km in both 2009 and 2012.  Since our ability to identify the end of the tail is limited by signal-to-noise considerations, the measured $\ell$ constitutes only a lower limit to the true length.  

First, we estimate the dust properties from these measurements.  The length and rise-time of the tail, $\tau \sim$ 1 day,  imply an average speed $V$ = $\ell$/$\tau \sim$ 3 km s$^{-1}$, which can be produced at a constant acceleration $a$ = 2$\ell /\tau^{2} \sim$ 0.07 m s$^{-2}$. For comparison, the solar gravitational acceleration at 0.14 AU is $g_{\odot}$ = 0.3 m s$^{-2}$, giving a ratio $\beta = a / g_{\odot} \sim$ 0.2.  The ratio of accelerations for a particle moving under the action of radiation pressure can  be written in terms of particle properties as

\begin{equation}
\beta = \frac{3 Q_{pr} F_{\odot} R_1^2}{4 G M_{\odot} c \rho r}.
\label{beta}
\end{equation}

\noindent Here, $Q_{pr}$ is the dimensionless radiation pressure factor, $F_{\odot}$ = 1360 W m$^{-2}$ is the Solar constant, $R_1$ = 1.5$\times$10$^{11}$ m is the number of meters in 1 AU, $G$ = 6.6$\times$10$^{-11}$ N kg$^{-2}$ m$^2$ is the gravitational constant, $M_{\odot}$ = 2$\times$10$^{30}$ kg is the mass of the Sun and $c$ = 3$\times$10$^8$ m s$^{-1}$ is the speed of light.   Particle quantities $\rho$ and $r$ are the density and radius, respectively.  We assume $Q_{pr}$ = 1, $\rho$ = 3000 kg m$^{-3}$ and substitute $\beta$ = 0.2 into Equation (\ref{beta}) to obtain $r \sim$ 1 $\mu m$.   In other words, the sudden emergence of the long tail implies a large $\beta$, corresponding to particles about 1 $\mu m$ in radius.  Particles of this size have scattering parameter $x = 2 \pi r / \lambda \sim 9$ at the $\lambda$ = 0.7 $\mu m$ wavelength of observation, near the peak scattering efficiency \citep{1983asls.book.....B}.  Presumably, much smaller particles exist but contribute weakly to the effective cross-section because they are inefficient scatterers ($x \ll$ 1) while larger particles ($x \gg$ 1) may exist and scatter efficiently, but are relatively rare and slow-moving (and would be confined to the vicinity of the nucleus in our data).

\begin{figure}
\epsscale{1.0}
\begin{center}
\plotone{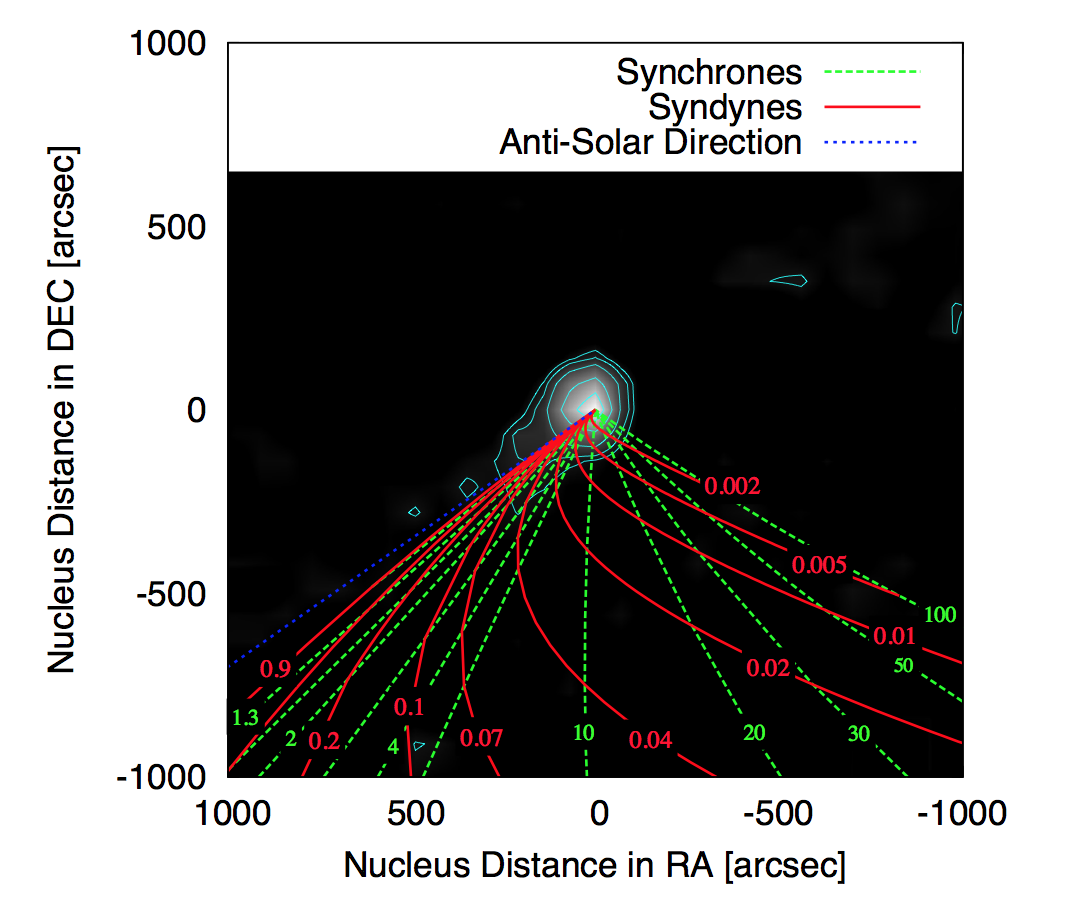}
\caption{Phaethon on UT 2009 June 20, 06:49 compared with dust models.  Synchrones (green) correspond to ejection at 100, 50, 30, 20, 10, 5.3, 4.3, 3.3, 2.3, 1.8, 1.3, 0.8 days before the date of the image. Syndynes (red) correspond to $\beta$ = 0.002, 0.005, 0.01, 0.02. 0.04, 0.07, 0.1, 0.2, 0.4, 0.9, as marked.   \label{jessica}
} 
\end{center} 
\end{figure}

We next computed syndyne (particles with a single $\beta$ ejected over a range of times) and synchrone (particles with a range of $\beta$ ejected at one time) models of the trajectories of dust particles.  The models take account of orbital motion and projection into the plane of the sky as viewed from STEREO-A.  In making these trajectory calculations we assume that the particles are ejected from Phaethon with negligible initial velocity. Model results for 2009 June are plotted in Figure (\ref{jessica}). Equivalent calculations for 2012 May give comparable results and are not shown.   The syndynes for large particles (small $\beta$) would occupy a tail having a position angle inconsistent with the one observed, as would synchrones for particles older than a few days (see Figure \ref{jessica}).  The latter result is further consistent with the onset of the anomalous perihelion brightening within a day of our first tail detection.  The position angle of the tail and the sudden growth of the tail are both consistent with the action of radiation pressure on small particles.   

Estimates of the dust mass from the photometry taken in 2009 and 2012 \citep{2013AJ....145..154L}, give $M_d$ = 2.5$\times$10$^5 r_{\mu m}$ and $M_d$ = 4$\times$10$^5 r_{\mu m}$, respectively, where $r_{\mu m}$ is the particle radius expressed in microns.  We take the average, $M_d \sim$ 3$\times$10$^5$ $r_{\mu m}$, and substitute $r_{\mu m}$ = 1 to estimate the mass of efficient scatterers in the tail as $M_d \sim$ 3$\times$10$^5$ kg.   This mass is miniscule compared to the estimated nucleus mass (2$\times$10$^{14}$ kg) and Geminid stream mass \citep[10$^{12}$ to 10$^{13}$ kg;][]{1989MNRAS.240...73H, 1994A&A...287..990J}.    If ejected uniformly over $\tau \sim$ 1 day, the average mass loss rate from Phaethon would be $dM/dt \sim M_d /\tau \sim$ 3 kg s$^{-1}$.  At this rate, the timescale for replenishment of  the Geminid stream mass ($\sim$10$^{12}$ to 10$^{13}$ s or 2$\times$10$^4$ to 2$\times$10$^5$ yr) is far longer than the $\sim$10$^3$ yr dynamical stream age \citep{1989A&A...225..533G, 2007MNRAS.375.1371R}.   It thus seems unlikely that the particles contributing to the optical tail in Figures (\ref{img2009}) and (\ref{img2012}) are sufficient to supply the Geminid meteoroid stream. 

This conclusion is reinforced by the syndyne/synchrone models, which further show that particles with $\beta >$ 0.07, like those dominating the optical tail, are gravitationally unbound to the solar system.  Such particles cannot be a significant source of Geminid stream meteoroids. On the other hand, much larger, slower, potentially mass-dominant particles could exist while contributing little to the scattering cross-section at optical wavelengths.  Such particles would be subject to smaller acceleration by radiation pressure (c.f. Equation \ref{beta}) and would remain in the unresolved vicinity of the nucleus in our data. Furthermore, we see no reason to assume that  mass loss, even at perihelion, should occur in a steady state.  It is entirely possible, for instance, that the perihelion mass loss rate varies stochastically (perhaps by orders of magnitude) from orbit to orbit, analogous to the way in which steady erosion of a coastal headland by ocean waves leads to rare but mass-dominant landslides.  Thus,  we can conclude that the inferred mass loss rate is too small to supply the Geminids in steady state, but we cannot use the new data to rule-out the possibility that Phaethon continues to actively supply its own meteoroid stream.

\clearpage

\section{Summary}

We have discovered a tail on Geminid-parent asteroid (3200) Phaethon at perihelion. The tail unambiguously establishes the presence of on-going mass-loss, confirming our prior inferences based on unresolved photometry alone.  The key features and conclusions from this tail are:

\begin{enumerate}


\item The tail grows to full length ($\gtrsim$ 250,000 km) within a single day, implying acceleration from the nucleus at 0.07 m s$^{-2}$ or greater (about 0.2 times the local solar gravitational acceleration).  This large acceleration is consistent with the action of radiation pressure on spherical dust grains $\sim$ 1 $\mu m$ in radius.  

\item Taken together, the photometry and the inferred grain size indicate a tail mass $\sim$3$\times$10$^5$ kg and a mass production rate $\sim$3 kg s$^{-1}$.

\item Most particles in the optical tail follow gravitationally unbound orbits and thus do not contribute to the Geminid meteoroid stream.  Much larger, slower, potentially mass-dominant  and gravitationally-bound particles could be simultaneously ejected from Phaethon but would escape detection in our data.

\item Previously suggested mechanisms of thermal fracture and desiccation cracking of hydrated minerals remain plausible sources of Phaethon's tail.

\end{enumerate}

\acknowledgments
We thank Toshi Kasuga and Pedro Lacerda for reading the paper and Michael A'Hearn for his review.  This work was supported by a grant to DCJ from NASA's Planetary Astronomy program. The Heliospheric Imager instrument was developed by a collaboration that included the University of Birmingham and the Rutherford Appleton Laboratory, both in the UK, the Centre Spatial de Liege (CSL), Belgium, and the U.S. Naval Research Laboratory (NRL), Washington DC, USA. The STEREO/SECCHI project is an international collaboration.

\clearpage

\clearpage

\begin{deluxetable}{rcccccc}
\tablecaption{Observing Geometry and Tail Position Angles
\label{positionangles}}
\tablewidth{0pt}
\tablehead{
\colhead{UT Date and Time\tablenotemark{a}}  & \colhead{$R$\tablenotemark{b}}  & \colhead{$\Delta$\tablenotemark{c}} & \colhead{$\alpha$\tablenotemark{d}}   & \colhead{$\theta_{Ph}$\tablenotemark{e}} 
& \colhead{$\theta_{\odot}$[deg]\tablenotemark{f}}   }
\startdata

2009 June 19d 06h 49m & 0.147 & 1.027  & 57 & --- & 131  \\
20d 06h 49m & 0.140 & 0.972  & 79 & 128$\pm$23 & 124  \\
21d 06h 49m & 0.146 & 0.916  & 102 & 108$\pm$16 & 116  \\

2012 May 01d 08h 09m & 0.146 & 1.052  & 46 & --- & 137  \\
 02d 08h 09m & 0.140 & 1.003  & 67 & 124$\pm$17 & 126  \\
 03d 08h 09m & 0.146 & 0.950  & 88 & 110$\pm$18 & 117  \\

\enddata

\tablenotetext{a}{Start time of the image composite from which the tail properties were measured.  Each composite consists of images taken over a period of one day with corrections for rotation of the projected antisolar direction applied.  Other parameters in the table all refer to the start time.}
\tablenotetext{b}{Heliocentric distance, in AU}
\tablenotetext{c}{Phaethon to STEREO A distance, in AU}
\tablenotetext{d}{Phase angle, in degrees}
\tablenotetext{e}{Measured position angle of the tail and estimated 1-$\sigma$ uncertainty, in degrees}
\tablenotetext{f}{Position angle of the projected anti-Solar direction, in degrees}

\end{deluxetable}

\clearpage

\clearpage

\clearpage

\clearpage

\end{document}